%% file: tesseract.tex
\ifx\compilefullpaper\undefined
\documentclass[10pt, twocolumn, aps, nofootinbib, longbibliography, nobibnotes, superscriptaddress]{revtex4-1} 
\input{header}

\usepackage[utf8]{inputenc}
\usepackage[table]{xcolor}

\usepackage[frozencache,cachedir=.]{minted} 
\usepackage{makecell} 

\usepackage[only,llbracket,rrbracket]{stmaryrd}

\usepackage{enumitem} 
\usepackage{color}
\definecolor{cardinal}{rgb}{0.827, 0, 0}

\begin{document}

\fi

\vfuzz2pt 

\title{
Demonstration of quantum computation and error correction 
with a tesseract 
code 
}

\affiliation{Microsoft Azure Quantum}
\affiliation{Quantinuum}

\author{Ben W. Reichardt}
\affiliation{Microsoft Azure Quantum}
\affiliation{University of Southern California}

\author{David Aasen}
\author{Rui Chao}
\affiliation{Microsoft Azure Quantum}

\author{Alex Chernoguzov}
\affiliation{Quantinuum}

\author{Wim van Dam}
\affiliation{Microsoft Azure Quantum}

\author{John P. Gaebler}
\author{Dan Gresh}
\author{Dominic Lucchetti}
\author{Michael Mills}
\author{Steven A. Moses}
\author{Brian Neyenhuis}
\affiliation{Quantinuum}

\author{Adam Paetznick}
\author{Andres Paz}
\affiliation{Microsoft Azure Quantum}

\author{Peter E. Siegfried}
\affiliation{Quantinuum}

\author{Marcus P. da Silva}
\author{Krysta M. Svore}
\author{Zhenghan Wang}
\author{Matt Zanner}
\affiliation{Microsoft Azure Quantum}

\begin{abstract}
A critical milestone for quantum computers is to demonstrate fault-tolerant computation that outperforms computation on physical qubits.  
The tesseract subsystem color code protects four logical qubits in $16$ physical qubits, to distance four.  
Using the tesseract code on Quantinuum's trapped-ion quantum computers, we prepare high-fidelity encoded graph states on up to $12$ logical qubits, beneficially combining for the first time fault-tolerant error correction and computation.  
We also protect encoded states through up to five rounds of error correction.  
Using performant quantum software and hardware together allows moderate-depth logical quantum circuits to have an order of magnitude less error than the equivalent unencoded circuits.  
\end{abstract}

\maketitle

\section{Introduction}

Even the best qubits are insufficiently reliable to run large quantum algorithms.  By encoding qubits into error-correcting codes, dramatic reductions in effective error rates should be possible.  
Fault tolerance allows software to handle hardware flaws---with a cost in overhead.  

\begin{figure}[b]
{\raisebox{0cm}{\includegraphics[width=.3\textwidth]{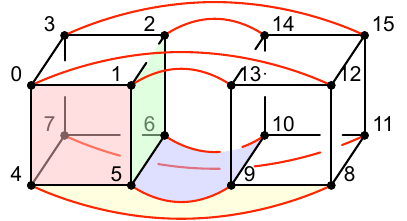}}}
\caption{
The $[[16,6,4]]$ color code on the 4D hypercube, or tesseract. 
Each of the $16$ vertices is a qubit.  Cubes are $X$ and $Z$ stabilizers, and squares are logical operators, e.g., 0145.  
}
\label{f:tesseract}
\end{figure}

We introduce a new fault-tolerance scheme that is well-suited to state-of-the-art quantum computers.  
The scheme comprises a $16$-qubit ``tesseract" code, sketched in \figref{f:tesseract}, and methods for error correction and encoded computation.  It is qubit efficient, with just a four-to-one overhead in qubit count, and also offers expeditious error correction (single shot, with only two extra qubits) and enough protection to reduce error rates substantially.  The code is distance four, 
so three physical faults have to occur to cause a logical error.  

The 
tesseract code has been studied before~\cite{delfosse2020short, PrabhuReichardt24code16}.  Our main innovation to the code is to deliberately sacrifice two of the original six encoded qubits.  Only having to protect four of the encoded qubits dramatically simplifies fault-tolerant error correction and computation.  
In particular, it lets error correction work by measuring operators supported on only four qubits, instead of eight, which improves efficiency and gives the single-shot property.  

We evaluate and demonstrate our fault-tolerance scheme on Quantinuum's H1 (20 qubit) and H2 (56 qubit) trapped-ion quantum computers~\cite{Moses23quantinuum}, obtaining and computing with up to $12$ logical qubits
in three code blocks.  
\tabref{f:experiments_summary} summarizes the experiments.  
The Path-$4$ experiment demonstrates an encoded, targeted CNOT gate within a code block; operations targeted to specific qubits allow more flexible computation than if we were limited to transversal gates in which every encoded qubit gets the same operation.  
Cube-$8$ uses three rounds of transversal CNOT gates between blocks, with two rounds of error correction; this demonstrates a deeper circuit, on more encoded qubits, for which good error correction 
during the computation 
is critical.  
Cat-$12$ prepares a high fidelity cat state on $12$ logical qubits.  
Finally, we show five rounds of error correction on four and eight encoded qubits, paving the way for deeper logical circuits on more encoded qubits.  All of these encoded circuits work with significantly lower error rates than the unencoded baseline versions.  

\begin{table}
\caption{\label{f:experiments_summary}
Computing fault tolerantly on encoded data gives dramatic error rate improvements over baseline, unencoded circuits.  Path-$4$ is run on H1, the others on H2.  
The first three experiments prepare graph states, on four, eight, or 12 qubits.  
Their reported error rates are averaged over $X$ and $Z$ measurement settings.  
Full data is given in \tabref{f:fulldata}.  
}
\setlength{\tabcolsep}{1.9pt}
\begin{tabular}{c@{$\!\!\!\!\!\!\!\!\!\!$}cc@{$\,\,\,\,$}c@{$\,\,\,\,$}c@{$\,\,\,\,$}c}
\Xhline{2\arrayrulewidth}
& && Baseline & Encoded & 
 \\[-.1cm]
\multicolumn{2}{c}{Experiment} & $\!\!\!$Qubits$\!$
& error rate & error rate & Gain \\
\hline
{$\;\;$\includegraphics[scale=.1]{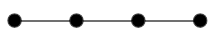}} & Path-$4$ & 4 & 
1.5(2)\% 
& $0.10^{+0.11}_{-0.06}\%$ & $15 \times$ \\
{$\;\;$\raisebox{-.1cm}{\includegraphics[scale=.1]{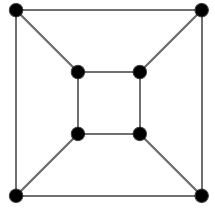}}} & Cube-$8$ & 8 & 
2.3(3)\% 
& $0.2^{+0.2}_{-0.1}\%$ & $11 \times$ \\
$\;\;$\raisebox{.05cm}{\scalebox{.5}{$\ket{0^{12}} + \ket{1^{12}}$}} & Cat-$12$ & 12 & 
2.4(3)\% 
& $0.11^{+0.16}_{-0.08}\%$ & $22 \times$ \\[.1 cm]
\hline \\[-.35cm]
\multicolumn{2}{c}{Error correction $5 \times$}
& \begin{tabular}{c}4 \\ 8\end{tabular} & \begin{tabular}{c}$2.7(4)\%$ \\ $5.6(6)\%$ \end{tabular} & \begin{tabular}{c} $0.11^{+0.21}_{-0.09}\%$ \\ $0.7^{+0.7}_{-0.4}\%$ \end{tabular} &\begin{tabular}{c}$24 \times$ \\ $8 \times$ 
\end{tabular} \\
\Xhline{2\arrayrulewidth}
\end{tabular}
\end{table}

Note that with a distance-four code, two faults may bring the system to an uncorrectable state.  When an uncorrectable state is detected, our fault-tolerance procedure rejects the trial.  This ``postselection" 
means that more trials are needed to collect a given amount of encoded data.  The acceptance rates in our experiments are at least $50\%$, so this extra time overhead is not critical.  It is far better than with a distance-two code for which a single fault can cause rejection.  Still, the overhead will get worse for longer experiments on more code blocks.  

Compared to codes that have been implemented previously, the tesseract code has a higher rate (ratio of encoded to physical qubits) and higher distance.  
These parameters come with tradeoffs.  A higher rate usually means a code requiring more physical qubits, better connectivity between those qubits, and more complex logical computation.  A higher distance often forces a lower rate, and sometimes more complicated fault-tolerance schemes, meaning physical error rates have to be lower before the code is useful.  For current ion trap devices, with 20+ well-connected and high-fidelity qubits, the tesseract code is in a sweet spot for achieving fault-tolerant computation.  

\medskip
\noindent
\textbf{Related work.}
The latest generations of quantum hardware are enabling impressive demonstrations of fault-tolerant quantum error correction and computation, validating fault tolerance theory's applicability to real devices.  

Distance-two codes have been implemented with two~\cite{GuptaIBM24magiccz}, three~\cite{MenendezRayVasmer23colorcode832, Wang23color832addition}, four~\cite{Yamamoto23phaseestimattionerrordetection}, eight~\cite{self22icebergcode}, and $48$ logical qubits, divided across $16$ $[[8,3,2]]$ code blocks~\cite{BluvsteinHarvard23neutralatoms}.  Distance-two codes are challenging to scale because they cannot correct errors, only detect them, and this leads to high rejection rates.  

For repeated error correction, Ryan-Anderson et al.~\cite{honeywell21steane} have experimented with up to six rounds of error correction, but on only one encoded qubit, in the $[[7,1,3]]$ color code, and after five rounds the logical error rate was over $10\%$.  Postler et al.~\cite{Postler23steaneec} implemented repeated error correction on the same code, and after one round the error rate was already about $35\%$.  
Krinner et al.~\cite{Krinner21repeatedsurfaceec} have implemented up to $16$ cycles of syndrome extraction on the $[[9,1,3]]$ surface code, equivalent to about $16/3 \approx 5.3$ rounds of error correction, with over $3\%$ logical error rate per cycle.  
Google Quantum AI and collaborators have implemented up to $250$ cycles of syndrome extraction on the $[[49,1,7]]$ surface code, equivalent to about $250 / 7 \approx 35.7$ rounds of error correction, with a $0.143(3)\%$ logical error rate per cycle~\cite{google23surface, google24surfacecode}.  Importantly, this makes the one logical qubit have a longer lifetime than a physical qubit in their system.  
Repeated error correction on multiple qubits has been implemented by Silva et al.~\cite{Silva24microsoft12qubitcode}.  They run three rounds of error correction on two qubits encoded in a $[[12,2,4]]$ code, with a logical error rate of only $0.8(3)\%$, also beating a physical baseline.  
Here we achieve five rounds of error correction, on up to eight encoded qubits, with lower logical error rates---up to $24 \times$ lower error per logical qubit per error correction round---and higher acceptance rates (\tabref{f:experiments_summary}).  

Yamamoto et al.~\cite{Yamamoto23phaseestimattionerrordetection} have combined computation with error detection.  
As for computation with error-correcting codes, Ryan-Anderson et al.~\cite{quantinuum22h1} study logical CNOT gates between two code blocks, for the $[[5,1,3]]$ and $[[7,1,3]]$ codes.  For their $[[5,1,3]]$ code experiment, some error correction is necessary for fault tolerance, and yet they find that it increases logical error rates.  For their $[[7,1,3]]$ code experiment, error correction is unnecessary, and so again they find that it increases logical error rates.  
Postler et al.~\cite{Postler21universal} and Mayer et al.~\cite{Mayer24steaneftexperiment} do a computation on two and three encoded qubits, respectively, using the $[[7,1,3]]$ code, notably including non-Clifford gates, 
but not using rounds of quantum error correction.  
%
We extend these capabilities by demonstrating a beneficial {combination} of fault-tolerant error correction with computation, perhaps for the first time.  

Creating large cat states, also known as GHZ states, is a common metric to advertise hardware progress.  For example, Quantinuum researchers have prepared on H2 the $20$-qubit cat state with an $86\%$ fidelity, and the $32$-qubit cat state with an $82\%$ fidelity~\cite{Moses23quantinuum}.  Using superconducting qubits, Bao et al.~\cite{Bao24catstate60} have prepared a $60$-qubit cat state with a $59\%$ fidelity.  (For a list of cat state experiments, see~\cite{Krenn24catstates}.)  
Encoded cat states can have much higher fidelities.  
Hong et al.~\cite{Hong24cat4onh2} prepare a four-qubit cat state encoded in a $[[25,4,3]]$ code on H2, with fidelity at least $99.5^{+0.2}_{-0.4}\%$.  
Bluvstein et al.~\cite{BluvsteinHarvard23neutralatoms} prepare a four-qubit cat state encoded in the $[[7,1,3]]$ code, with a fidelity of $72(2)\%$ using error correction, ranging to $99.85^{+0.1}_{-1.0}\%$ using full error detection.  We prepare an encoded cat state on $12$ logical qubits, with comparably high fidelity and a higher acceptance rate.  See \tabref{f:catstateexperiments}.  
(Note that our experiments, and \cite{Hong24cat4onh2}, measure $X$ and $Z$ error rates, $p_X$ and $p_Z$, with \tabref{f:experiments_summary} reporting $\tfrac12(p_X + p_Z)$.  The fidelity to the ideal cat state is between $1 - p_X - p_Z$ and $1 - \max (p_X, p_Z)$.)  

\begin{table}
\caption{\label{f:catstateexperiments}
Experiments preparing encoded cat states.  
}
\setlength{\tabcolsep}{1.9pt}
\begin{tabular}{cccc}
\Xhline{2\arrayrulewidth}
Reference & Logical qubits & Fidelity \\
\hline
\cite{Hong24cat4onh2} & 4 in $[[25,4,3]]$ code & $99.5^{+0.2}_{-0.4}\%$ to $99.7^{+0.2}_{-0.3}\%$ \\
\cite{BluvsteinHarvard23neutralatoms} & 4 in $[[7,1,3]]$ code & \scalebox{.9}{$\!\!\!\left\{\begin{array}{c} \\[.20cm] \end{array}\right.\!\!\!\!\!$}\begin{tabular}{c}72(2)\% error correction\\ $99.85^{+0.1}_{-1.0}\%$ error detection\end{tabular}$\!\!\!$ \\
This work & 12 in $[[16,4,4]]$ code & $99.82^{+0.12}_{-0.4}\%$ to $99.90^{+0.1}_{-0.3}\%$ \\
\Xhline{2\arrayrulewidth}
\end{tabular}
\end{table}

\medskip
\noindent
\textbf{Our experiments.}
In general, our work is notable in showing multiple rounds of error correction on multiple code blocks, with more encoded qubits and a variety of logical operations---all at error rates an order of magnitude lower than the unencoded versions.  

The experiments demonstrate different capabilities of the tesseract code fault-tolerance scheme.  
The first three experiments are to prepare and verify encoded graph states~\cite{Hein06graphstates}.  
Preparing graph states reliably is a good test of entangling Clifford gates. 


\smallskip
The Path-$4$ experiment demonstrates an encoded CNOT gate between two logical qubits in the same code block.  For codes encoding more than one qubit per block, targeted operations within the block are usually much more difficult than applying the same operation to every encoded qubit.  A common technique, from~\cite{Gottesman97}, is to teleport the encoded qubits of interest into their own code blocks, that are otherwise empty, so that they can be addressed separately---but this wastes the high rate capability of the code.  Targeted internal operations without overhead is an important advantage of the tesseract scheme.  

Here and in the other experiments, we use the basic tesseract code fault-tolerance ingredients from \secref{s:tesseractcode} as separate modules, plugging them together to get the desired encoded circuit.  Had we simply wanted to prepare the encoded graph state with as high fidelity as possible, it is likely that a specially tailored encoding circuit could perform better.  We felt it was more important to demonstrate that the fault-tolerance modules worked well, since they can be used for a broad variety of experiments and circuit constructions.  

The unencoded Path-$4$ experiment uses three CNOT gates, while the encoded experiment gets away with one.  
Two of the encoded CNOT gates come basically for free, by permuting the physical qubits. 
Permuting qubits is an easy operation for the ion trap hardware.  

\begin{figure}
$\!\!\!\!\!$ \includegraphics[scale=1]{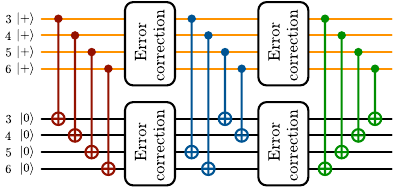} $\;\;$ \raisebox{1cm}{\includegraphics[scale=.55]{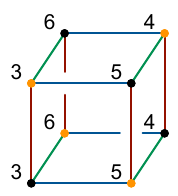}}
\caption{Encoded circuit to prepare the cube graph state, in two code blocks.} \label{f:encodedcube}
\end{figure}

\smallskip
The Cube-$8$ experiment shows a deeper logical circuit, on more encoded qubits, with essential error correction.  
Preparing the cube graph state is a challenge because it requires more CNOT gates ($12$) than any other bipartite eight-qubit graph state~\cite{CabelloDanielsenLopezTarridaPortillo10graphstatedatabase}.  
The encoded circuit involves three rounds of transversal CNOT gates, between two code blocks, with two rounds of error correction on each code block (\figref{f:encodedcube}).  The error correction is key.  Without it, single faults can cause logical errors---e.g., an $X$ error on the control code block could spread to a weight-three $X$ error on the target block---and in simulations the circuit is immediately overwhelmed by noise.  
%
We do not know of prior work 
that combines fault-tolerant error correction and computation to reduce logical error rates.  

\smallskip
The Cat-$12$ experiment is not as deep, but shows even more entangled logical qubits, with a high fidelity.  

\smallskip
The goal of our repeated error correction experiments is to better protect more encoded qubits, through more error correction rounds.  This is a challenge because in the future, as the field moves toward large-scale fault-tolerant computations, experiments will deploy many more logical qubits through much deeper logical circuits.  As periodic error correction is needed to maintain fault tolerance~\cite{AharonovBenOr99, KnillLaflammeZurekScience98, AliferisGottesmanPreskill05}, highly reliable repeated error correction is a prerequisite.  
Although \cite{Silva24microsoft12qubitcode} and this work both show that repeated error correction gives an improvement compared to physical baselines, more important is the low absolute error rate.  

\section{Tesseract subsystem color code} \label{s:tesseractcode}

The $[[16,6,4]]$ tesseract code is a self-dual color code on the 4D hypercube, encoding into $16$ physical qubits six logical qubits protected to distance four 
(\figref{f:tesseract}).  This code was introduced in~\cite{delfosse2020short} and simulated extensively, with a fault-tolerance scheme requiring weight-eight measurements, in~\cite{PrabhuReichardt24code16}.  
The main idea behind our scheme is to sacrifice two encoded qubits, not using them to store data (except sometimes temporarily).  
The sacrificed qubits can be thought of as ``gauge qubits"~\cite{Bacon05operator}, even though they have the same distance as the data qubits.  
The resulting $[[16,4,4]]$ code we call the tesseract subsystem code.  

\begin{figure}
{\includegraphics[width=.3\textwidth]{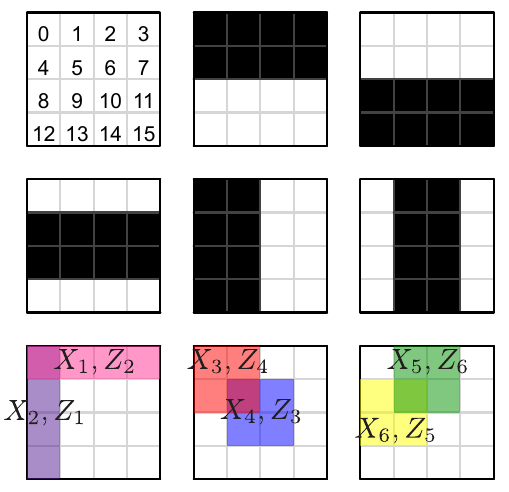}}
\caption{
It is convenient to arrange the tesseract code's $16$ qubits in a $4 \times 4$ grid.  The $X$ and $Z$ stabilizers are supported on pairs of rows and pairs of columns; a set of generators is shown in black.  Last, we highlight the supports of the logical operators in the basis we use.  For example, logical $Z_2$ is $Z_0 Z_1 Z_2 Z_3$.  Observe that the weight-four logical operators are not self dual, but come in three pairs of two.  
}
\label{f:stabilizers}
\end{figure}

The advantage of sacrificing two encoded qubits is not hard to see.  Consider the code presentation in \figref{f:stabilizers}.  Notice that $X^{\otimes 4}$ along any row of the grid is a representative of encoded $X_1$.  The product of any two $X^{\otimes 4}$ rows is a stabilizer.  Similarly, $X^{\otimes 4}$ along any column is a representative of encoded $X_2$.  The product of any two columns gives a stabilizer.  Thus measuring the first two logical qubits, using weight-four measurements, is enough to determine the syndromes for all $X$ stabilizers, enabling $Z$ error correction.  Weight-four measurements are much easier to make fault tolerantly to distance four than weight-eight measurements~\cite{prabhu21, PrabhuReichardt24code16}.  
Moreover, it is fault tolerant to measure the four rows in parallel followed by the columns in parallel; error correction is ``single shot"~\cite{delfosse2020short}.  

\begin{figure}
\subfigure[\label{}]{\includegraphics[scale=.8]{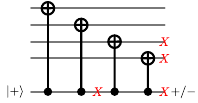}}$\qquad$
\subfigure[\label{f:measurement_fullyft}]{\includegraphics[scale=.8]{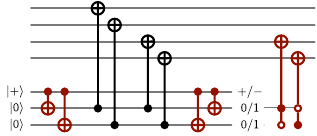}}
\subfigure[\label{f:measurement_oneflag}]{\includegraphics[scale=.8]{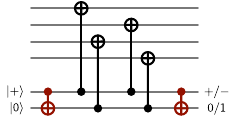}}$\qquad$
\subfigure[\label{f:measurement_xxxxandzzzz}]{\includegraphics[scale=.8]{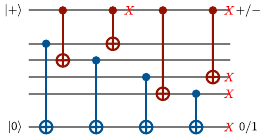}}
\caption{Circuits to measure weight-four operators.
The circuit in~(a) measures $X^{\otimes 4}$, but it is not fault tolerant, as a single $X$ fault (red) on the ancilla qubit can propagate to a weight-two error on the data.  
The circuit in~(b), using one syndrome and two flag qubits, is fully fault tolerant.  The corrections can be tracked classically, as part of the ``Pauli frame."
(c)~This circuit's single flag qubit is enough to detect a possible correlated error, but not correct it.  A subsequent measurement will have to take the flag into account to correct the answer.  
(d) We can also measure $X^{\otimes 4}$ and $Z^{\otimes 4}$ simultaneously.  In our applications, the results should be uniformly random.  But repeating the $X^{\otimes 4}$ and $Z^{\otimes 4}$ four times, on disjoint qubit sets (\figref{f:ftmeasurements}), if, e.g., one of the $Z^{\otimes 4}$ measurements disagrees with the other three, that can mean that either a weight-one $X$ error has been detected, or a weight-two correlated $X$ error has spread to the data.}
\label{f:weight4measurements}
\end{figure}

Even better, when measuring $X^{\otimes 4}$ across a row (logical $X_1$), we can simultaneously measure $Z^{\otimes 4}$ across the same row (logical $Z_2$).  As shown in \figref{f:measurement_xxxxandzzzz}, $X^{\otimes 4}$ and $Z^{\otimes 4}$ can be efficiently and fault-tolerantly measured in parallel, with one measurement outcome flagging the other for possible correlated errors~\cite{Reichardt18steane}.  Each pair of weight-four measurements takes only eight CNOT gates, with two ancilla qubits.  We will reuse the same two ancilla qubits for all measurements on a code block, so one code block uses $18$ physical qubits in hardware.  Interestingly, this method allows for simultaneously correcting $X$ and $Z$ errors, a major efficiency advantage over having sequential procedures as is common.  

Finally, 
the two gauge qubits give workspace for implementing targeted operations, such as H, CNOT or CZ, on any one or two encoded qubits within the same or different code blocks.  Essentially, these operations can be implemented by teleporting through the gauge qubits.  

\smallskip

\begin{figure}
\subfigure[\label{}]{\includegraphics[scale=.5]{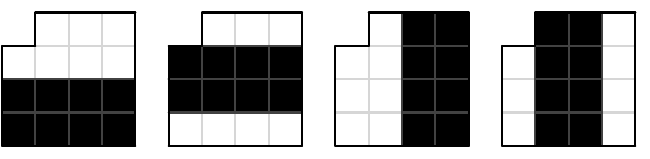}}
\subfigure[\label{f:832colorcode}]{\includegraphics[scale=.6]{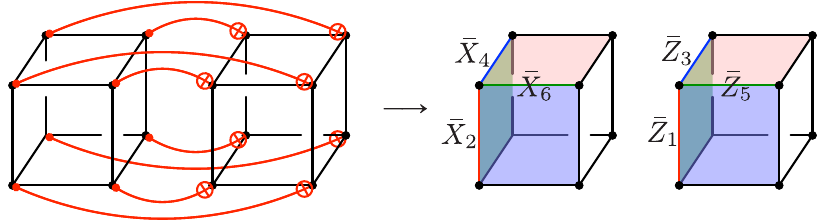}}
\subfigure[\label{}]{\includegraphics[scale=.6]{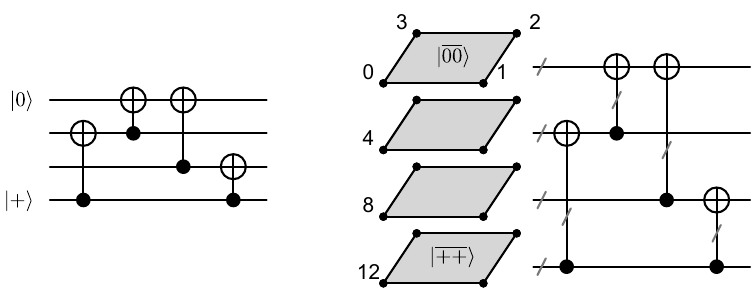}}
\caption{Relationships to other codes.  (a) Removing a qubit from the $[[16,6,4]]$ code leaves the well-known $[[15,7,3]]$ Hamming code.  (b) Applying CNOT gates between two halves of the 16-qubit code disentangles it into two $[[8,3,2]]$ color codes on the cube; the left color code has $X$ distance $2$ and $Z$ distance $4$, and the right color code vice versa.  (c) Left: an encoding circuit for the $[[4,2,2]]$ color code.  Stacking four copies of this color code, with the first in encoded $\ket{00}$ (a cat state) and the last in encoded $\ket{{+}{+}}$, and applying the same encoding circuit transversally yields our $[[16,4,4]]$ code with fixed gauge qubits~\cite{BreuckmannBurton22foldtransversalclifford}.}
\label{f:code_relationships}
\end{figure}

The tesseract code is closely connected to other codes (\figref{f:code_relationships}).  Notably, it can be obtained by applying transversal CNOTs between the $[[8,3,2]]$ color code on the 3D cube~\cite{Campbell16eight32colorcode}, and its dual; or from four copies of the $[[4,2,2]]$ color code on the square.  

\medskip
\noindent
\textbf{Fault-tolerant error correction.}
An efficient fault-tolerant error correction procedure is the foundation for any fault-tolerant computation scheme.  One round of error correction consists of measuring $X^{\otimes 4}$ and $Z^{\otimes 4}$ across four rows, then down four columns.  The $X^{\otimes 4}$ measurement outcomes across the rows may be random, but should be correlated.  In the absence of noise, one should obtain either $(0,0,0,0)$ or $(1,1,1,1)$.  If one outcome disagrees with the other three, it indicates a $Z$ error in that row, e.g., $(0,1,0,0)$ and $(1,0,1,1)$ both indicate a $Z$ error in the second row.  The column $X^{\otimes 4}$ measurements should similarly identify the column of the $Z$ error, so it can be corrected.  If two rows disagree with the others, e.g., $(0,0,1,1)$ or $(0,1,0,1)$, this indicates an uncorrectable error and we reject the trial.  

Actually, it is not so simple.  In the circuit of \figref{f:measurement_xxxxandzzzz}, a single $Z$ fault when measuring a column can cause a weight-two $Z$ error in that column of qubits.  We cannot reject these first-order events or the rejection rate would be too high; worse, if a single fault can cause weight-two errors, the logical error rate will only be second order.  For a distance-four code, the logical error rate should be third order.  

Fortunately, when a single $Z$ fault causes a weight-two $Z$ error in a column, it will be flagged by the $X^{\otimes 4}$ column measurement.  Thus when measuring $X^{\otimes 4}$ along the rows, we proceed differently when a column has been flagged for $Z$ errors.  If a flag has been raised, then we accept $(0,0,1,1)$ or $(1,1,0,0)$ row measurement outcomes, and correct them by applying $ZZII$ (or equivalently $IIZZ$) down the column.  

\begin{figure}
{\includegraphics[scale=.4]{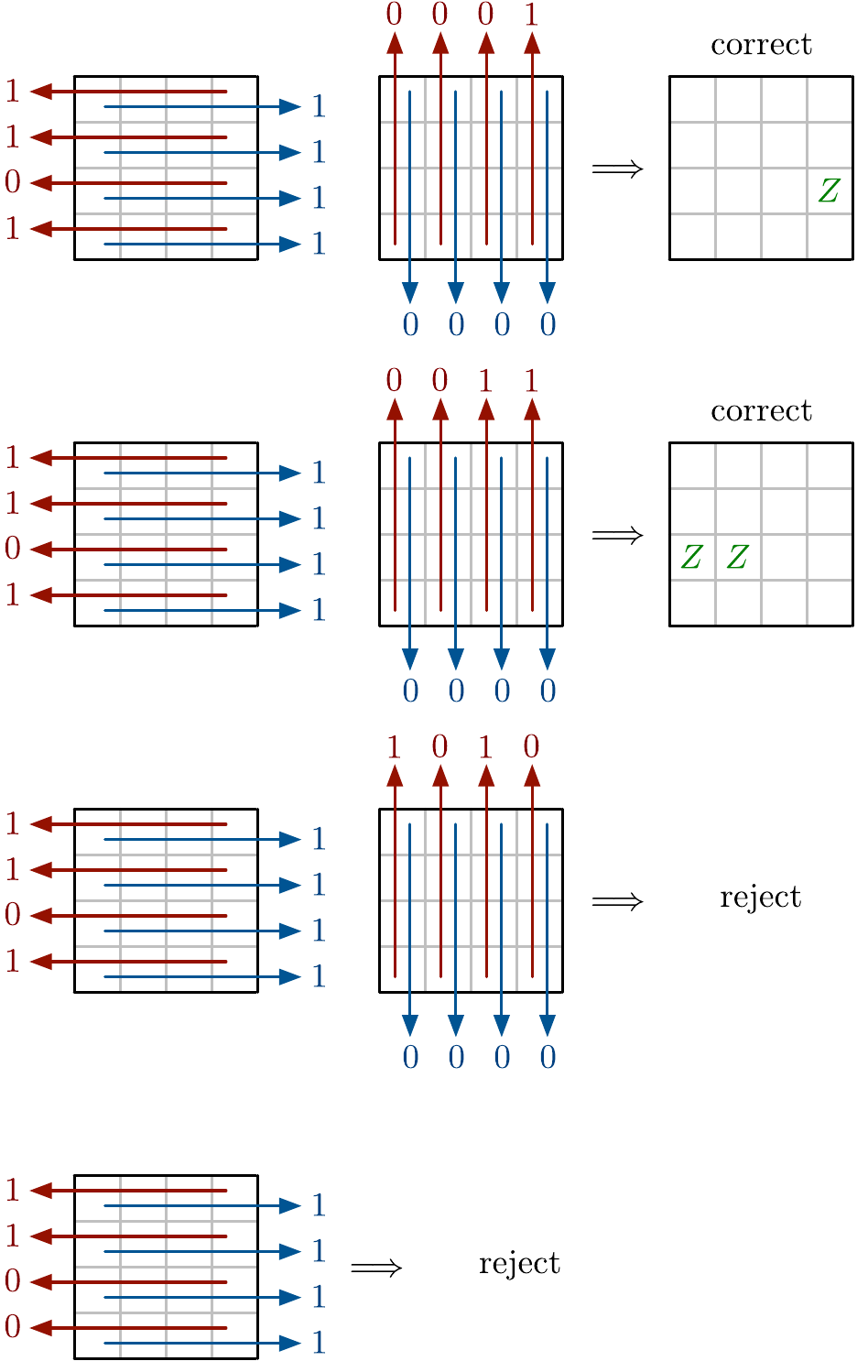}}
\caption{Error correction examples.  
$X^{\otimes 4}$ outcomes are red, $Z^{\otimes 4}$ outcomes are blue, and corrections are indicated in green.  
If one row $X$ measurement disagrees with the others, and one column $X$ measurement disagrees with the others, then we apply a $Z$ correction to the row, column intersection.  A disagreeing row $X$ measurement also flags that row for a possible correlated $ZZII$ or $IIZZ$ error.  However, if there are two $0$ and two $1$ row measurements (with no column flagged in the previous column measurements), then we have to reject.}
\label{f:errorcorrectionexamples}
\end{figure}

\begin{figure*}
{\includegraphics[scale=.4]{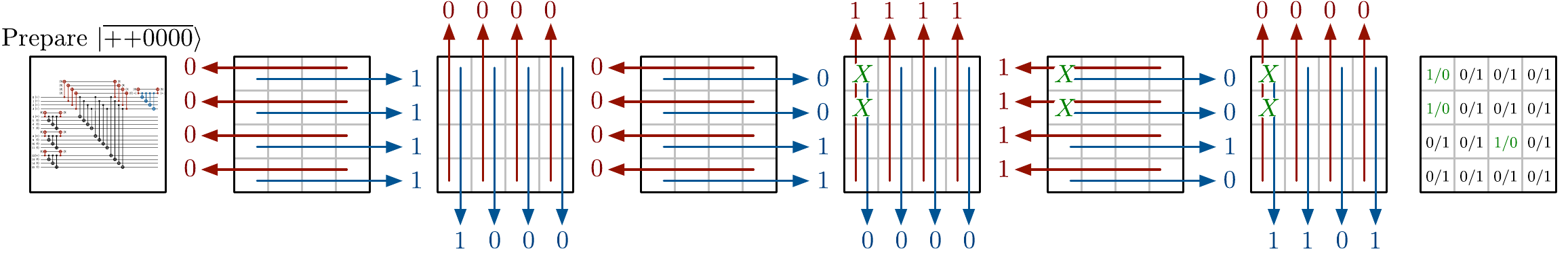}}
\caption{Repeated error correction experiment with three rounds of $X$ and $Z$ error correction.  Starting from the left, each round of error correction consists of measuring $X^{\otimes 4}$ and $Z^{\otimes 4}$ across four rows, then down four columns, using the circuit in \figref{f:measurement_xxxxandzzzz}.  
The error correction rules are ``rolling," meaning that if an error is first detected in the column measurements, then the subsequent row measurements are used to correct it.}
\label{f:repeatederrorcorrection}
\end{figure*}

Figure~\ref{f:errorcorrectionexamples} illustrates a variety of cases, and \figref{f:errorcorrectionrules} lays out the error correction rules in detail.  
Figure~\ref{f:repeatederrorcorrection} shows a complete example of three rounds of repeated error correction, with a flagged $Z^{\otimes 4}$ measurement leading to an $XXII$ correction applied to that column after the next row measurements.  

\begin{figure}
\begin{minted}{python}
if flagX == -1: # no row flagged already
  if sum(measX) == 2: 
    return "reject"
  if sum(measX) in (1,3): 
    if sum(measX) == 1: flagX = measX.index(1)
    else:               flagX = measX.index(0)
else: # row flagX in (0,1,2,3) flagged
  if sum(measX) in (1,3): 
    if sum(measX) == 1: col = measX.index(1)
    else:               col = measX.index(0)
    frameZ[4*flagX + col] += 1  # Z correction
  if sum(measX) == 2: 
    if measX in ([0,0,1,1], [1,1,0,0]): 
      frameZ[[4*flagX, 4*flagX+1]] += 1  # ZZII
    else: 
      return "reject"
  flagX = -1
\end{minted}
\caption{Error correction rules.  
This code processes $Z$ error correction for column measurements.  
Let {\tt measX} store the results of $X^{\otimes 4}$ measurements on four columns.  Let $\text{\tt{flagX}} = -1$ if no preceding row $X^{\otimes 4}$ measurement was flagged, or $\text{\tt flagX} \in \{0,1,2,3\}$ the flagged row with a possible weight-one $Z$ error or correlated $ZZII$ or $IIZZ$ error.  
A $Z$ correction is stored in the Pauli frame.  
Similar code works for $X$ error correction, and for row $X$ and $Z$ measurements.
} 
\label{f:errorcorrectionrules}
\end{figure}

It is not obvious that the resulting scheme is fault tolerant.  A full argument requires case checking, the tricky case being when there are two faults, e.g., a flagged correlated error and one additional fault.  This should not cause a logical error.  


\medskip
\noindent
\textbf{Fault-tolerant operations.}
There is a large variety of primitives enabling fault-tolerant computation with the tesseract subsystem code.  
We will mention here only those essential for our experiments.  

\begin{figure}
\subfigure[\label{f:preparation_xxzzzz}]{\includegraphics[scale=.8]{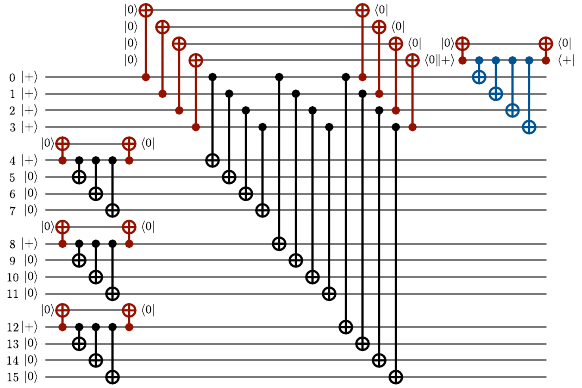}}
\subfigure[\label{f:preparation_xzxzxz}]{\includegraphics[scale=.8]{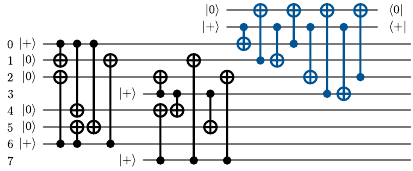}}
\caption{Initialization circuits.  (a) Fault-tolerant postselected circuit to prepare encoded $\ket{{+}{+}0000}$.  The red gates are for flags, and the blue gates for measuring a stabilizer.  
(b) Taking two copies of this postselected circuit fault tolerantly prepares encoded $\ket{{+}0{+}0{+}0}$.  Each copy is $\ket{000}$ encoded in the $[[8,3,2]]$ color code.  
}
\label{f:initialization}
\end{figure}

The experiments all begin by initializing basic encoded states $\ket{{+}{+}0000}$ or $\ket{{+}0{+}0{+}0}$, with the circuits in \figref{f:initialization}.  Here the flag and syndrome measurements are postselected, meaning that if any measurement is nontrivial we reject the state and start again.  Rejections during state preparation are in theory less of a concern than rejections deeper into a computation, when starting over is costly.  In the full experimental data below, we break out as ``prerejection rate" the fraction of trials that were rejected during state preparation.  

Of course, transversal $X$ and $Z$ measurements, followed by classical decoding, can be used to destructively measure all of the logical $X$ or $Z$ operators fault tolerantly.  Interestingly, there is also a simple procedure for measuring half the logical qubits in the $X$ basis and half in the $Z$ basis.  
For example, to measure $Z, X, Z, X, Z, X$, apply row-transversal CNOT gates from row 1 to row 4, and row 2 to row 3; then measure each qubit in the top half in the $X$ basis, and each in the bottom in the $Z$ basis.  As shown in \figref{f:832colorcode}, the CNOT gates divide the logical qubits between two $[[8,3,2]]$ color codes.  Although these codes only have distance two, they have distance four in the direction that matters ($Z$ distance four for the top half, $X$ distance four for the bottom), so one can reliably decode the measurement results.  
We will use this measurement procedure in the repeated error correction experiments below.  

\begin{figure}
{\includegraphics[scale=.3]{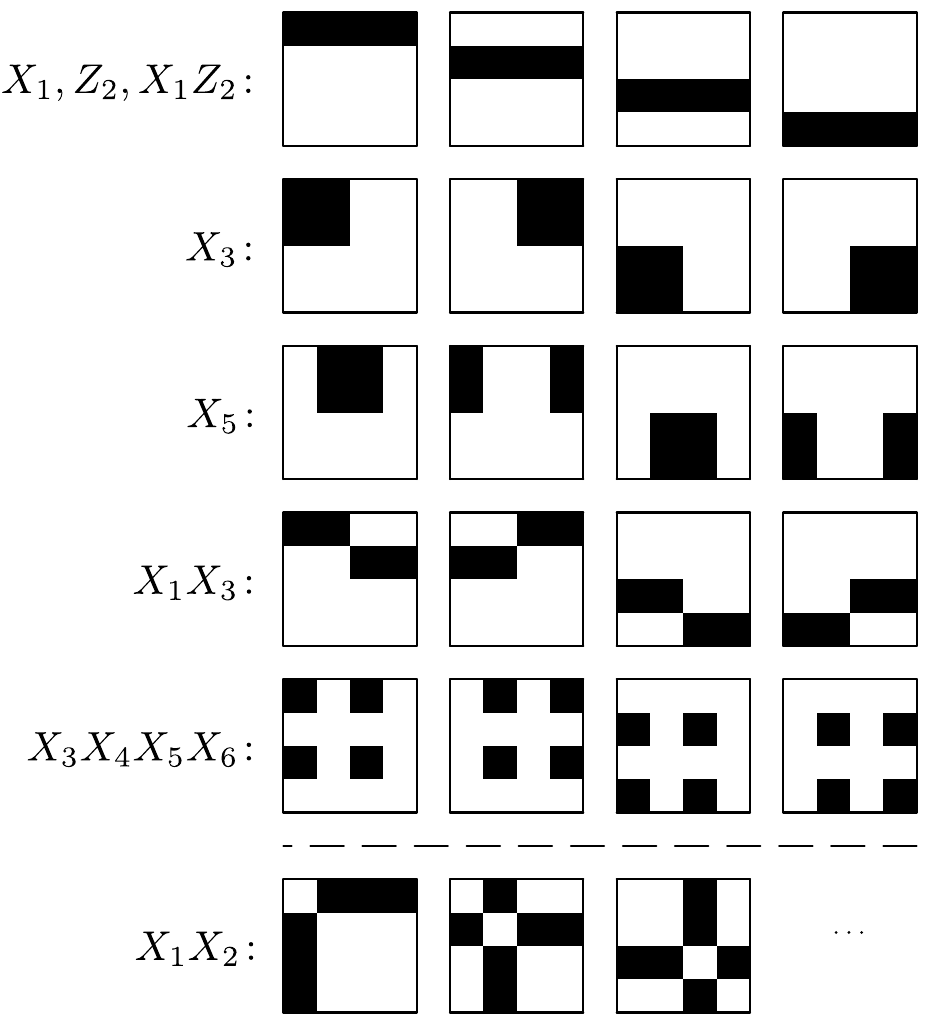}}
\caption{Logical measurements.  Each logical operator $X_1$, $Z_1, \ldots, Z_6$ has four qubit-disjoint weight-four representatives.  Measuring the representatives fault tolerantly and taking the majority of the results, postselecting on no tie, fault tolerantly measures that logical operator.  Many weight-two logical operators can also be measured this way, as well as higher-weight logical operators.  Importantly, $X$ or $Z$ logical measurements supported on both qubits of a pair (1,2), (3,4) or (5,6) can \emph{not} generally be measured this way, as their minimum-weight representatives can have weight six.  But, e.g., $X_1 Z_2$ has four $Y^{\otimes 4}$ representatives.  Shown is a selection of logical operators and their minimum-weight representatives' supports.}
\label{f:ftmeasurements}
\end{figure}

Importantly, we can also projectively measure single logical qubits, and some operators across multiple logical qubits, with single-shot weight-four measurements.  (For most codes, this would require multiple repeated rounds of measurements~\cite{delfosse2020short}.)  The error-correction procedure above uses this property to measure logical qubits 1 and~2.  Figure~\ref{f:ftmeasurements} shows more examples.  

In particular, these projective logical measurements allow for measurement-based computation~\cite{RaussendorfBriegel01cluster}, using the gauge qubits $1$ and~$2$ as workspace.  For example, to get an encoded CNOT gate from logical qubit~$i$ to~$j$, start with, say, logical qubit~$2$ in $\ket 0$.  Measure $X_2 X_i$ (correcting with $Z_i$ if the result is $1$), then measure $Z_2 Z_j$ (correcting with $X_2 X_i$ if the result is $1$), and finally measure $X_2$ (correcting with $Z_2 Z_j$ if the result is~$1$).  
The weight-four measurements can be made with either of the circuits in Figs.~\ref{f:measurement_fullyft} or~\ref{f:measurement_oneflag}; we use the latter, more efficient circuit, and if a flag is raised the next measurement (in the dual basis) can correct the possible correlated error.  

\smallskip

\begin{figure}
\setlength{\tabcolsep}{1.9pt}
\begin{tabular}{c@{$\qquad$}c}
Permutation & Logical effect \\
\hline \\[-.1in]
\raisebox{-.5cm}{\includegraphics[scale=.5]{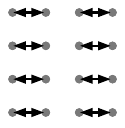}} & $e$ \\[.2in]
\raisebox{-.5cm}{\hspace{-.08cm}\includegraphics[scale=.5]{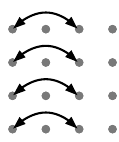}} & (35)(46) \\[.2in]
\raisebox{-.5cm}{\hspace{0cm}\includegraphics[scale=.5]{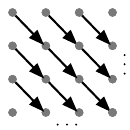}} & (34)(56) \\[.2in]
\raisebox{-.5cm}{\hspace{-.06cm}\includegraphics[scale=.5]{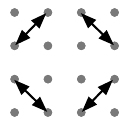}} & (15)(26) \\[.2in]
\raisebox{-.5cm}{\hspace{-.04cm}\includegraphics[scale=.5]{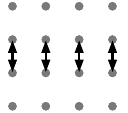}} & 
\raisebox{-.75cm}{\includegraphics[scale=.5]{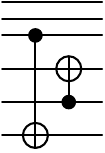}} \\[.2in]
\end{tabular}
\caption{Permutation automorphisms.  There are $16 \cdot 20160$ qubit permutations that preserve the code.  For example, applying any of the four permutations $e,(12)(34),(13)(24),(14)(23)$ to the columns and another to the rows (16 possibilities) 
has trivial logical effect.  Shown is a set of group generators, and their logical effects.  Note that certain combinations of logical CNOT gates can be implemented with permutations.  
}
\label{f:permutation_automorphisms}
\end{figure}

Many qubit permutations preserve the tesseract code space, and they can have a nontrivial logical effect.  Figure~\ref{f:permutation_automorphisms} gives the full group of permutation automorphisms.  

\section{Experiments} \label{s:experiments}

The subsections below give the details for the encoded and unencoded experiments.  
Experimental data 
is collected in \tabref{f:fulldata}.  Certainly 
more trials would be useful to  
narrow the confidence intervals.  However, it is already clear that the encoded circuits have significantly lower error rates than the unencoded baselines.  

\subsection{Quantum computers H1 and H2} 

The experiments described in this section were performed on the Quantinuum H1 and H2 trapped-ion quantum computers, which have 20 and 56 physical qubits, respectively, with all-to-all connectivity through ion-transport operations~\cite{quantinuum22h1, Moses23quantinuum, DeCross24randomcircuitsh2}.  Since the time of the referenced publications improvements were made to the H2 two-qubit gate laser, and the transport and cooling times were further optimized resulting in the following error parameters: two-qubit gate error $1.15(5) \times 10^{-3}$, single-qubit gate error $2.9(4) \times 10^{-5}$, SPAM error $1.47(9) \times 10^{-3}$, and memory error per qubit per depth-one circuit time $2.5(5) \times 10^{-4}$, where the depth-one circuit time is $\sim\!68$ ms.  
Current specifications for the H1 device can be found at~\cite{quantinuum24h1}.  

Note that using a distance-four code makes the logical error rate particularly sensitive to physical error parameters, with a cubic dependence.  
 
The specific structure of the quantum error correction circuit described in this work can leave some qubits idling for times much longer than the depth-one circuit time, which most accurately describes timing for circuits with densely packed two-qubit gates and arbitrary connectivity.  The naive approach originally taken by the compiler's qubit routing algorithm focused on maximizing the number of gates that can be done in parallel and was found to be suboptimal for these circuits.  Trading some parallelism for faster rearrangement resulted in reduction of overall circuit execution duration and consequently smaller overall memory error.  Although the improvements from these compiler optimizations are extremely circuit dependent we find for these circuits a 10--15\% reduction in total execution time.  
 
On H2, to further reduce memory error due to slow drifts in the qubit frequency for qubits with long idle times we insert dynamical decoupling pulses.  After the quantum circuit has been compiled to a set of transport instructions and gate operations, single-qubit gates are added to perform dynamical decoupling on qubits with more than 300ms between gates.  These dynamical decoupling pulses are inserted opportunistically as the qubits move through the gating zones during the transport operations used to reconfigure the qubits between rounds of two-qubit gates.  

\begin{table*}
\caption{\label{f:fulldata}
Experimental data.  
A trial is ``prerejected" if it is rejected during the initial state preparation, of encoded $\ket{00{+}{+}{+}{+}}$, $\ket{{+}{+}0000}$ or $\ket{{+}0{+}0{+}0}$.  Our preparation circuits allow prerejection to occur with first-order probability in the error rate; this is acceptable because only that block's initialization must be restarted and not the whole computation.  A trial is ``postrejected" if it is rejected any time after the initial state preparation, due to two faults being detected in close proximity.  With distance-four fault tolerance, postrejection should occur with a second-order probability.  
}
\setlength{\tabcolsep}{2pt}
\begin{tabular}{r@{$\quad$}ccccccc}
\Xhline{1.8\arrayrulewidth}
 & Meas. & & & & Acceptance & & \\[-.1cm]
Experiment & basis & Trials & Prerejected & Postrejected & rate & Errors & Error rate \\
\hline
Path-$4$ encoded
 & $X$ & 3000 & 220 & 89 & 90(1)\% & 3 & $0.12^{+0.18}_{-0.09}\%$ \\
 & $Z$ & 3000 & 192 & 96 & 90(1)\% & 2 & $0.08^{+0.16}_{-0.06}\%$ \\
 unencoded 
 & $X$ & 6000 & --- & --- & --- & 88 & $1.5(3)\%$ \\
 & $Z$ & 6000 & --- & --- & --- & 88 & $1.5(3)\%$\\
\hline
Cube-$8$ encoded
 & $X$ & 2000 & 298 & 272 & 71(2)\% & 2 & $0.2^{+0.3}_{-0.1}\%$ \\
  & $Z$ & 2000 & 256 & 238 & 75(2)\% & 3 & $0.2^{+0.3}_{-0.2}\%$  \\
unencoded 
 & $X$ & 6000 & --- & --- & --- & 151 & $2.5(4)\%$ \\
 & $Z$ & 6000 & --- & --- & --- & 119 & $2.0^{+0.4}_{-0.3}\%$ \\
\hline
Cat-$12$ encoded
 & $X$ & 2000 & 482 & 49 & $73(2)\%$ & 1 & $0.08^{+0.24}_{-0.07}\%$  \\
 & $Z$ & 2000 & 478 & 30 & 75(2)\% & 2 & $0.1^{+0.3}_{-0.1}\%$ \\
unencoded 
 & $X$ & 6000 & --- & --- & --- & 130 & $2.2^{+0.4}_{-0.3}\%$ \\
 & $Z$ & 6000 & --- & --- & --- & 163 & $2.7(4)\%$ \\
\hline
\hspace{-.15cm}\begin{tabular}{r}5 rounds of 4 qubit \\[-.1cm] error correction\end{tabular}
 & --- & 2500 & 297 & 148 & $82^{+1}_{-2}\%$ & 2 & $0.11^{+0.21}_{-0.09}\%$ \\
\begin{tabular}{r}teleportation \\[-.1cm] baseline
\end{tabular}
 & --- & 6000 & --- & --- & --- & 163 & $2.7(4)\%$ \\
\hline
\hspace{-.15cm}\begin{tabular}{r}5 rounds of 8 qubit \\[-.1cm] error correction\end{tabular}
 & --- & 2000 & 579 & 418 & 50(2)\% & 7 & $0.7^{+0.7}_{-0.4}\%$ \\
\begin{tabular}{r}teleportation \\[-.1cm] baseline
\end{tabular}
 & --- & 6000 & --- & --- & --- & 338 & $5.6(6)\%$ \\
\Xhline{1.8\arrayrulewidth}
\end{tabular}
\end{table*}

\subsection{Path-$4$ state preparation experiment}

The stabilizers for the $4$-qubit path graph state are $XXII, IXXX, ZZZI, IIZZ$.  The unencoded state preparation circuit is simply 
$$
\includegraphics[scale=1.2]{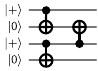}
$$
We follow this preparation with transversal $X$ or $Z$ measurements ($6000$ trials for each), and count the fraction of trials in which a stabilizer violation is observed.  

\medskip 

The encoded state preparation procedure has four steps: 
\begin{enumerate}[leftmargin=*]
\item 
Prepare encoded $\ket{{+}0{+}0{+}0}$, as in \figref{f:preparation_xzxzxz}.  
\item 
Permute the qubits, as in \figref{f:permutation_automorphisms}, to implement two encoded CNOT gates, and get two encoded Bell pairs, on qubits $3,6$ and $4,5$.  
\item 
Implement an encoded CNOT gate, from qubit~$6$ to~$5$.  This is done with three fault-tolerant measurements, using qubit~$2$ as workspace: 
\begin{enumerate}[leftmargin=*]
\item Measure $X_2 X_6$ (correcting $Z_2$ if the result is~$1$).  This is done by using the circuit in \figref{f:measurement_oneflag} to measure each of the $X^{\otimes 4}$ operators with supports 
$$
\includegraphics[scale=.3]{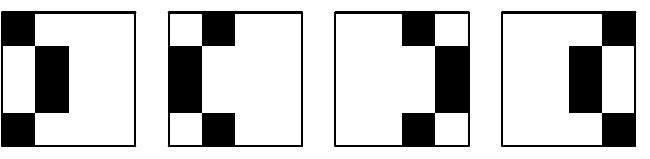}
$$
\item Measure $Z_2 Z_5$ (correcting $X_2 X_6$ if the result is~$1$).  This time measure the $Z^{\otimes 4}$ operators 
$$
\includegraphics[scale=.3]{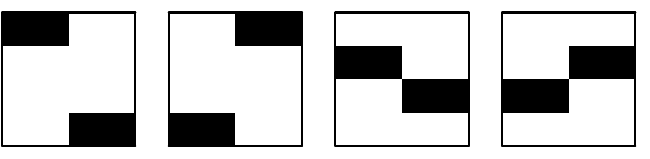}
$$
Note that each of these $Z$ operators overlaps each of the previous $X$ operators on exactly one qubit.  If one of the $X$ measurements was flagged, then the $Z$ measurements can correct the possible $X$ or $XX$ error.  A trial with two flags is rejected.  
\item Measure $X_2$ (correcting $Z_2 Z_5$ if the result is~$1$).  The $X^{\otimes 4}$ operators have support 
$$
\includegraphics[scale=.3]{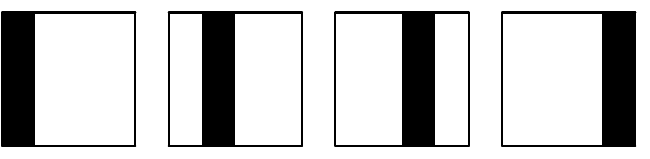}
$$
Once again, an $X$ or $XX$ error from a flagged $Z$ measurement can be corrected.  
\end{enumerate}
\item
Transversal $X$ or $Z$ measurement.  The measurement results are updated with the stored Pauli frame, then decoded classically, taking into account a possible $X$ measurement flag.  
\end{enumerate}

We ran this experiment on H1, with a total of 6000 trials divided between the $X$ and $Z$ measurement settings.  5403 trials were accepted, and, among those, five logical errors were found.  See \tabref{f:fulldata}.  

For H2, simulations suggest that the physical baseline should be about the same, but the encoded circuit's error rate perhaps about $10 \times$ lower.  Since it would take many trials to distinguish H2 from H1, we did not run the experiment on H2.  

In an earlier version of the experiment, we used the two-flag circuit of \figref{f:measurement_fullyft} to make the measurements.  That is slightly simpler, because flags don't have to be passed to the next step.  However, in H1 and H2 simulations the one-flag version has about a 5x lower logical error rate.

\subsection{Cube-$8$ state preparation experiment}

A circuit to prepare the 
$8$-qubit cube graph state is: 
$$
\includegraphics[scale=1.0]{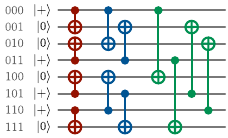}
\qquad
\raisebox{.4cm}{\includegraphics[scale=.5]{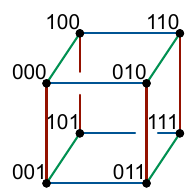}}
$$
Here, each qubit is labeled by its coordinates in a unit cube.  The first round of CNOT gates connects qubits in the $z$ direction, and the second and third rounds connect in the $y$ and $x$ directions.  
As the cube has $12$ edges, there are $12$ CNOT gates.  

The encoded procedure starts with encoded $\ket{00{+}{+}{+}{+}}$ and $\ket{{+}{+}0000}$ in two code blocks, prepared as in \figref{f:preparation_xxzzzz}.  As shown in \figref{f:encodedcube}, three rounds of transversal CNOT gates between the blocks generate the encoded cube state.  
In the second and third CNOT gate rounds, the target block's qubits are permuted so as to permute the encoded qubits.  (We swap two rows in the second round, and swap two columns in the third round.)  
The state is then measured transversally, and decoded.  

The four error correction steps are essential here, because CNOT gates copy errors between blocks.  Without error correction, for example, a single $X$ error on the first block could spread to a weight-three $X$ error on the target block, which would decode to a logical error.  (In fact, fault tolerance requires only $X$ error correction on the control block, and $Z$ error correction on the target block.  We choose to correct $X$ and $Z$ errors on both blocks.)  

Note that when interpreting the results, measurement flags also need to be passed between the blocks.  For example, if a column in the first block is flagged for a possible $XX$ error, since that error would be copied to the second block the flag also needs to be copied.  If the second block was already flagged for an $X$ or $XX$ error, then we reject.  Aside from passing flags like this, we do not use correlated error decoding between the blocks.

\subsection{Cat-$12$ state preparation experiment}

The $12$-qubit cat state $\tfrac{1}{\sqrt 2}\big(\ket{0^{12}} + \ket{1^{12}}\big)$ can be prepared with four rounds of CNOT gates, $11$ CNOTs total: 
$$
\includegraphics[scale=.8]{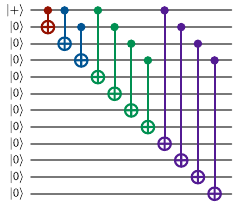}
$$
It is a graph state for the star graph.  In fact, in this case the physical baseline we compare to is the dual cat state $\tfrac{1}{\sqrt 2}\big(\ket{+^{12}} + \ket{-^{12}}\big)$, because it has slightly lower error rates in simulation.  

To prepare the encoded cat state, on three code blocks, we start by preparing an encoded cat state $\ket{0^4} + \ket{1^4}$ in qubits $3,4,5,6$ of one code block.  This is similar to the Path-$4$ experiment, but simpler: 
\begin{enumerate}[leftmargin=*]
\item 
Prepare encoded $\ket{{+}0{+}0{+}0}$, as in \figref{f:preparation_xzxzxz}.  
\item 
Permute the qubits, as in \figref{f:permutation_automorphisms}, to implement two encoded CNOT gates, and get two encoded Bell pairs, on qubits $3,6$ and $4,5$.  
\item 
Finally, merge the two Bell pairs by measuring $Z_4 Z_6$ (correcting $X_3 X_6$ if the result is~$1$).  This is done by using the one-flag circuit in \figref{f:measurement_oneflag} to measure each of the $Z^{\otimes 4}$ operators with supports 
$$
\includegraphics[scale=.3]{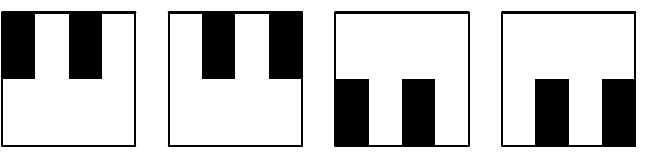}
$$
\end{enumerate}
Then apply transversal CNOT gates into two copies of $\ket{{+}{+}0000}$, and measure all $48$ physical qubits in the $X$ or $Z$ basis.  With transversal $X$ measurements, the classical decoder must take into account that one of the $Z$ measurements may have raised a flag, indicating a possible $ZZ$ error in the first code block.  

Although there are $12$ encoded qubits, using all $56$ physical qubits in H2, the encoded circuit is shallower and simpler than the Cube-$8$ experiment.  

\subsection{Five rounds of repeated error correction, with $10$ one-qubit teleportations}

Ryan-Anderson et al.~\cite{honeywell21steane} found that each round of error correction introduced about $2.7\%$ logical error to their one encoded qubit, biased toward logical $Z$ errors.  Silva et al.~\cite{Silva24microsoft12qubitcode} prepared $\ket{{+}{+}}$, a state sensitive to $Z$ errors, encoded in the $[[12,2,4]]$ code, and found that three rounds of error correction introduced $0.8(3)\%$ logical error.  Since $\ket{{+}{+}}$ is invariant under $X$ errors, the total error rate is slightly higher.  We prepare and protect encoded $\ket{{+}0{+}0}$.  This state is sensitive to both $X$ and $Z$ errors.  In fact, though, simulations starting with encoded $\ket{0000}$ and $\ket{{+}{+}{+}{+}}$ did not show significant differences in the logical error rate, suggesting that the logical $Z$ bias may have been largely ameliorated by the memory optimizations and dynamical decoupling introduced in~H2.  

\smallskip

The unencoded physical baseline for our repeated error correction experiment is the following circuit consisting of five rounds, having in each round two one-qubit teleportation steps: 
\begin{equation*} 
\includegraphics[scale=.8]{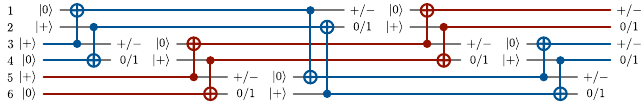}
\end{equation*}
The Pauli corrections for the one-qubit teleportations are not shown, but are $Z$ for $X$ measurements, and $X$ for $Z$ measurements.  Our interpretation for this circuit is that data qubits $3,4,5,6$ are cyclically rotating through the six qubit positions.  We call a trial a success if in the end, after Pauli corrections, the final four measurement outcomes are ${+}, 0, {+}, 0$, and otherwise declare an error.  

\begin{figure}
\subfigure[]{
\includegraphics[scale=.38]{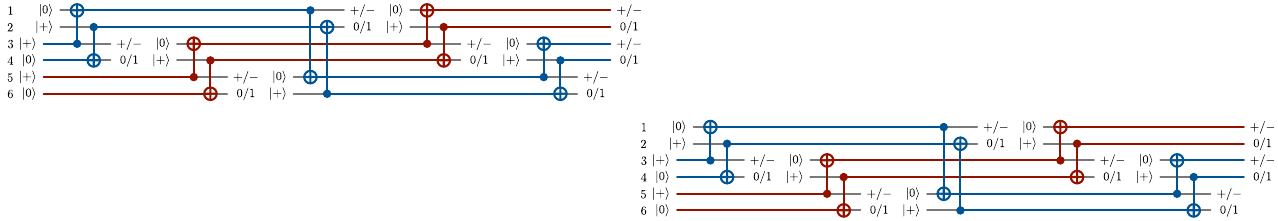}
}
\subfigure[\label{f:repeatederrorcorrectionbaseline_withbarrier}]{
\includegraphics[scale=.7]{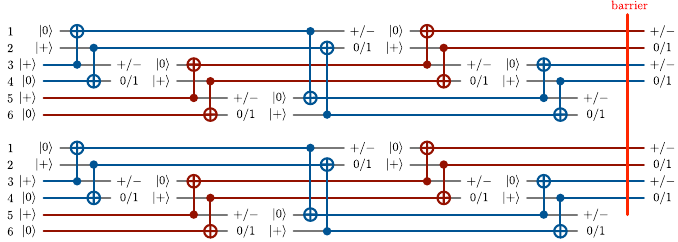}
}
\caption{
(a) Two independent circuits intended to run in parallel could instead be scheduled sequentially.  The compiler barrier in (b) disallows fully sequential compilation.  
}
\end{figure}

In the eight-qubit version of this experiment, we repeat the above circuit 
twice in parallel, as shown in \figref{f:repeatederrorcorrectionbaseline_withbarrier}.  We enforce that the circuits are run in parallel, and not scheduled sequentially, by inserting a compiler barrier on all eight data qubits before the final measurements.  This forces all operations before the barrier to finish before any operations after it begin~\cite{Quantinuum20userguide}.  
Sequential execution would presumably be easier since it lets the device devote its limited parallelism to one block at a time, and needs fewer qubits in memory.  

\smallskip

The encoded version 
is \emph{not} just repeated error correction with alternating row and column measurements, as in \figref{f:repeatederrorcorrection}.  The baseline comparison for that experiment would be four idling physical qubits, and  
idle ion trap qubits are a very good memory.  

\begin{figure*}
\subfigure[]{
\includegraphics[scale=.35]{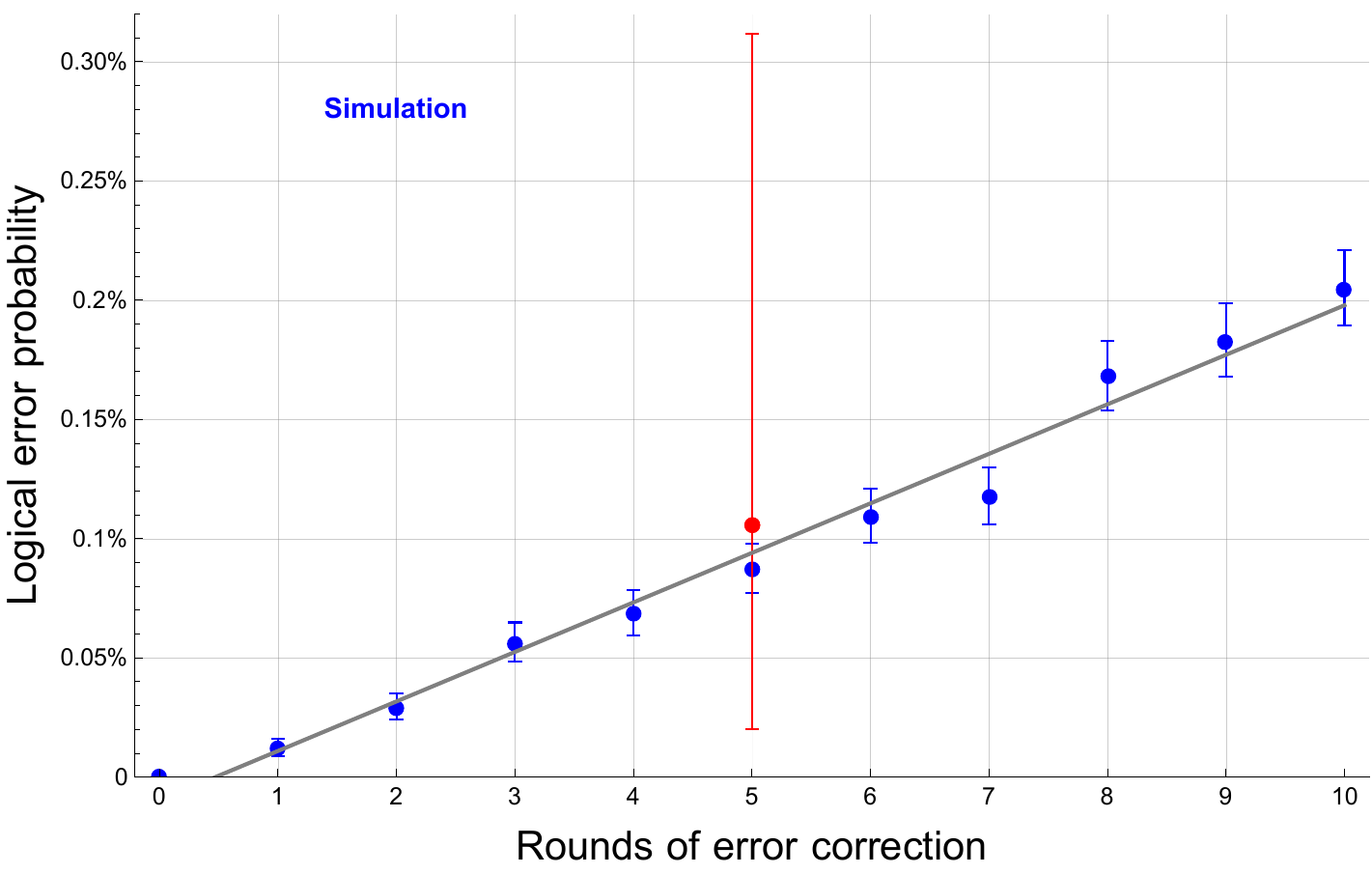}
$\qquad$
\includegraphics[scale=.35]{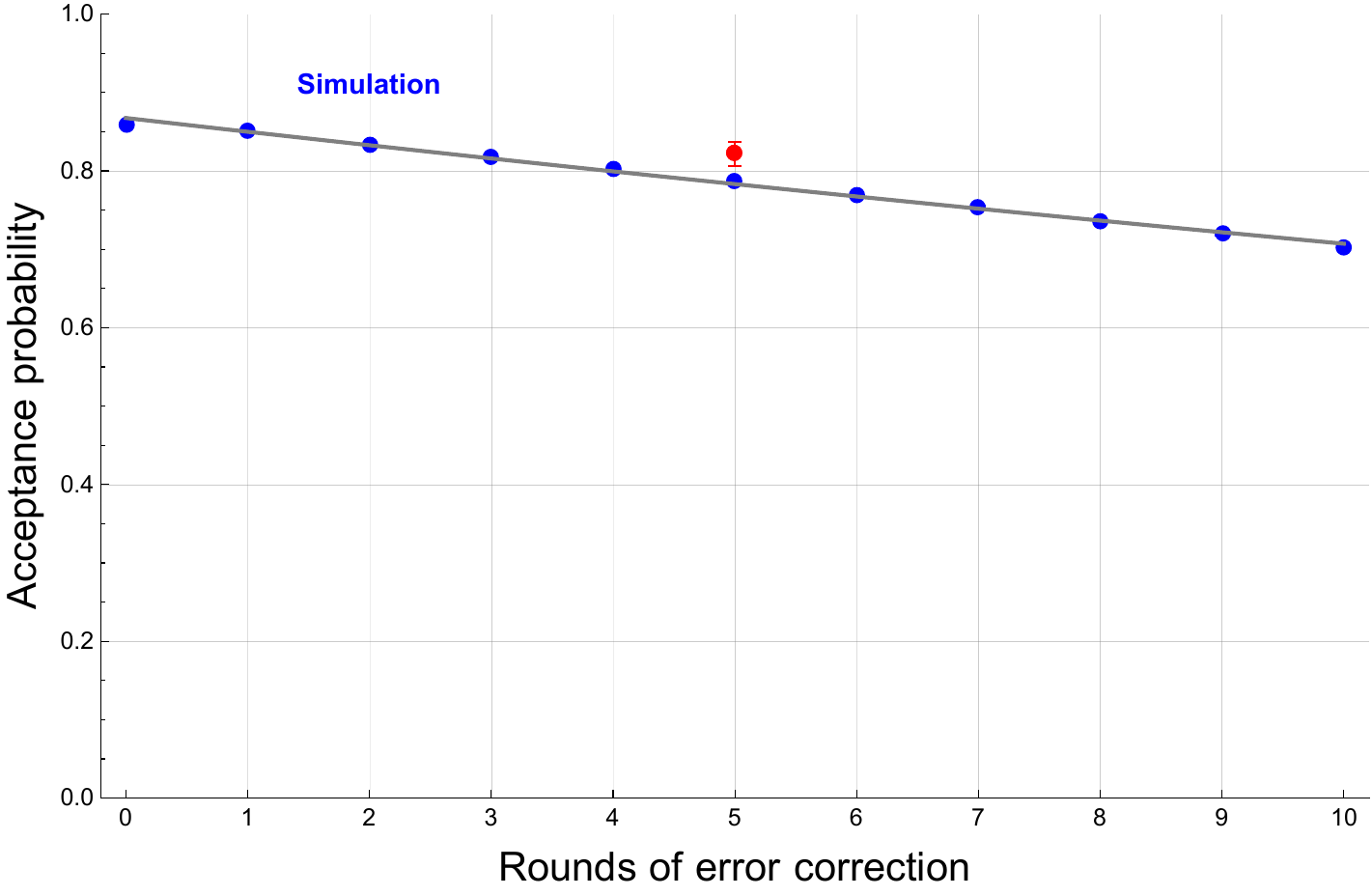}
}
\subfigure[]{
\includegraphics[scale=.35]{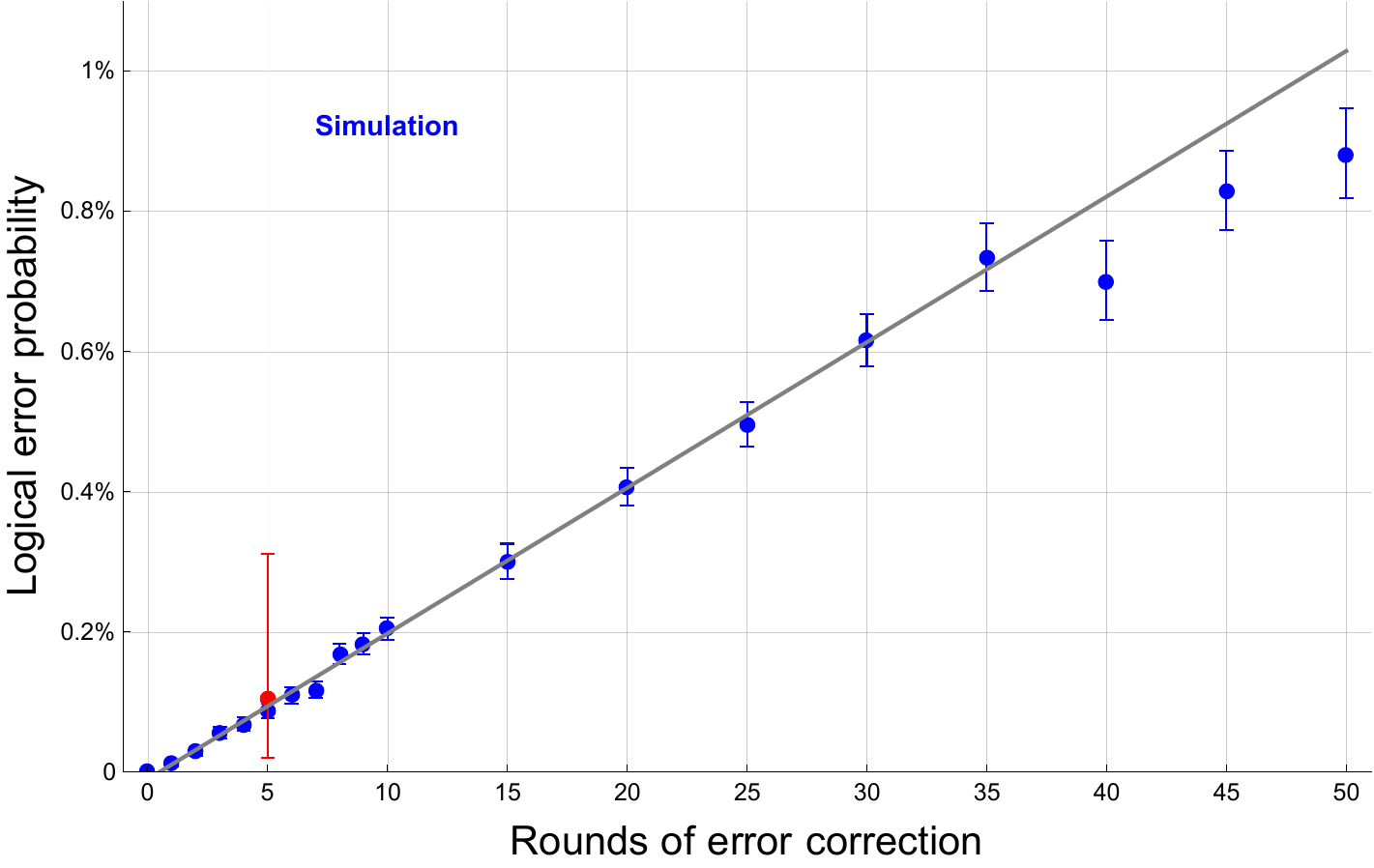}
$\qquad$
\includegraphics[scale=.35]{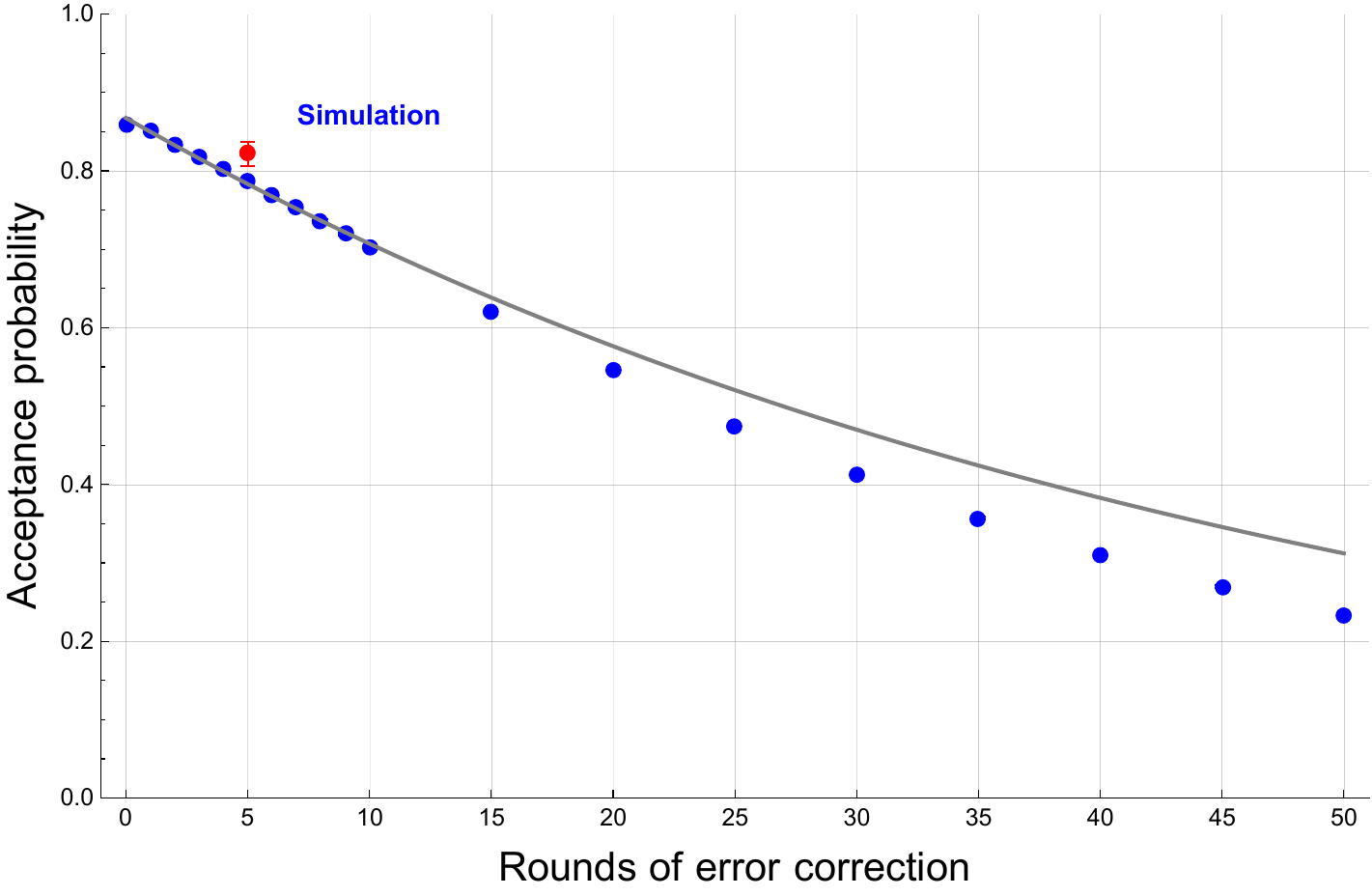}
}
\caption{(a) Simulated data showing the logical error probability and acceptance rate plotted versus rounds of error correction, on one code block.  The red points are real data from our five-round experiment.  (b) Simulations extended to 50 rounds of error correction.  The fit line and exponential are still based on the first 10 rounds.} \label{f:1blocksimulatedrotatedec}
\end{figure*}

Instead, we implement a more complicated version of repeated error correction that simultaneously makes the above cyclic one-qubit teleportations 
among the six encoded qubits.  This alternative error correction experiment proceeds as follows: 
\begin{enumerate}[leftmargin=*]
\item Prepare encoded $\ket{{+}0{+}0{+}0}$, as in \figref{f:preparation_xzxzxz}.  
\item Measure together encoded $Z_1$ and $X_2$ to initialize encoded $\ket{0{+}{+}0{+}0}$.  (If the $Z_1$ outcome is~$1$, correct with $X_1$.  If the $X_2$ outcome is~$1$, correct with $Z_2$.)  
\item Repeat five times: 
\begin{enumerate}[leftmargin=*]
\item Measure together $X_1 X_3$ (correction $Z_1$) and $Z_2 Z_4$ (correction $X_2$), using the circuit in \figref{f:measurement_xxxxandzzzz}.  
\item Rotate the encoded qubits cyclically forward two steps, i.e., with the permutation $(3,1,5)(4,2,6)$.  This logical permutation is implemented with the physical qubit permutation $(0,2,5)(3,6,4)(8,15,10)(9,12,14)$.  
\item Measure $Z_1$ (correction $X_1 X_5$) and $X_2$ (correction $Z_2 Z_6$) by measuring $X^{\otimes 4}$ and $Z^{\otimes 4}$ down each column.  This restores the gauge qubits to $\ket{0{+}}$.  
\end{enumerate}
In each step, the four $X^{\otimes 4}$ and $Z^{\otimes 4}$ measurement outcomes are used to correct errors, including possible flagged correlated errors, just as in \figref{f:repeatederrorcorrection}.  
\item Destructively measure encoded $X_3$, $Z_4$, $X_5$, $Z_6$.  As explained in \secref{s:tesseractcode}, and similar to \figref{f:832colorcode}, apply row-transversal CNOT gates from row 1 to row 2, and row 4 to row 3; then measure each control qubit in the $X$ basis, and each target qubit in the $Z$ basis.  Decoding the $[[8,3,2]]$ code for the control half gives $X_3$ and $X_5$, while decoding the target half gives $Z_4$ and $Z_6$.  
\end{enumerate}
In the repeated error correction experiment of \figref{f:repeatederrorcorrection}, the logical measurement outcomes do not matter, i.e., measurements $0000$ and $1111$ are treated the same.  The gauge qubits' states are irrelevant.  However, in this experiment the logical measurement outcomes do matter, since they determine the one-qubit teleportation Pauli corrections.  We expect this experiment to be slightly more challenging.  

\smallskip

A state-vector emulator for the H2 system is available from Quantinuum~\cite{Quantinuum24emulator}.  
In our experiments, the emulator gives very accurate results for small circuits such as our physical baseline comparisons on up to $12$ qubits.  It is less accurate for our experiments on one code block, tending to underestimate acceptance probabilities.  Nonetheless, we have simulated up to $50$ rounds of the one-qubit teleportation version of error correction, on one code block.  Figure~\ref{f:1blocksimulatedrotatedec} plots the simulated logical error and acceptance probabilities.  The logical error probability's growth is at least consistent with a straight line, as one would hope, while the acceptance probability has a slow exponential decline.  (The acceptance rate starts below~$1$ because about $13.6\%$ of trials are discarded due to a detected state preparation error.)  In the first $10$ error correction rounds, the logical error rate increases by $2.1(1) \times 10^{-4}$ per round, i.e., this is the slope of the fit line, and $2.02(2)\%$ of the surviving trials are discarded.  
Certainly we would have liked to validate these simulations with more experiments, but our time on H2 was limited.  


We have not run similar simulations for two code blocks, because the state-vector emulator cannot simulate the $36$ qubits this would require.  The stabilizer emulator can simulate all $56$ qubits in the H2 device, but while it can be useful for guidance it is not accurate enough in $18+$ qubit simulations to draw conclusions.

\section{Outlook}

We have presented a collection of fault-tolerance gadgets for the tesseract subsystem code, and have used them together to demonstrate a variety of fault-tolerant, error-corrected computations.  These computations beneficially combine fault-tolerant computation and error correction, for the first time, while outperforming the physical baselines in fidelity.  

It would be interesting to study 
the composability of the different gadgets, measuring the error rate through each extended rectangle~\cite{AliferisGottesmanPreskill05}, and verifying that logical error rates compose sub-additively.  
There are also more fault-tolerance gadgets to develop. 
We have not 
demonstrated a full set of Clifford operations.  In particular, we need targeted CNOT gates between blocks, e.g., from logical qubit~$1$ of block~$1$ to logical qubit~$2$ of block~$2$, and logical $S$ gates.  
Due to identities like 
$$
\includegraphics[scale=.9]{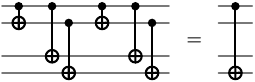}
$$
targeted CNOTs within a block, together with transversal CNOTs between them, imply targeted CNOTs between blocks---but that is an inefficient and inelegant construction.  
In fact, a targeted CNOT between blocks can be done using measurements, as in the Path-$4$ experiment, and a logical $S$ gate can be implemented 
with a carefully prepared ancilla code block.  We have not yet tested these gadgets experimentally.  

Beyond that, it is important to achieve full universality, potentially with multiple encoded non-Clifford gates.  There are several options for achieving logical universality to different degrees of fault tolerance: non-fault tolerant (distance one), distance two, or distance four.  We aim to develop some of these techniques and deploy them in a computation including both Clifford and non-Clifford operations.  

Finally, there are good prospects for scaling up the tesseract subsystem code fault-tolerance scheme to larger experiments.  The H2 device has limited parallelism, but, with dynamical decoupling, it exhibits much better memory than~H1.  This is why we ran the four-qubit 
error correction experiment on H2. 
Early experiments with five rounds of error correction on $12$ logical qubits---an admittedly difficult test---have shown a potentially problematically high error rate, $2^{+3}_{-1}\%$.  There is room for improvement, to increase both the number of logical qubits and the depth of logical computation, and to further reduce the logical error rate.  Improvements to the hardware, the compiler, and the fault-tolerance scheme all need to come together to address this challenge.

\section*{Acknowledgments}

We would like to thank Dennis Tom and Jenni Strabley for general feedback and support of the collaborative work.  
We also thank the Quantinuum hardware team for enabling these experiments.

\ifx\compilefullpaper\undefined  
\bibliographystyle{halpha-abbrv}
\bibliography{q}

\end{document}
\fi

%% file: header.tex
\usepackage{amsfonts, amssymb, amsmath, amsthm}
\usepackage{latexsym}
\usepackage[tracking=smallcaps]{microtype}	
\usepackage{url}
\usepackage{color}
\definecolor{DarkGray}{rgb}{0.1,0.1,0.5}
\definecolor{curgray}{rgb}{0.5,0.5,0.5}
\definecolor{curblue}{rgb}{0.04,0.11,0.64}
\definecolor{curpurple}{rgb}{0.65,0.16,0.58}
\definecolor{curorange}{rgb}{1,0.32,0}

\PassOptionsToPackage{hyphens}{url}

\usepackage[colorlinks=true,breaklinks=true, linkcolor=DarkGray,citecolor=DarkGray,urlcolor=DarkGray]{hyperref}	

\usepackage{graphicx}
\usepackage{subfigure}

\newcommand{\ket}[1]{{|#1\rangle}}

\newcounter{sprows}

\newlength{\spheight}
\newlength{\spraise}

\newlength{\commentslength}

\newcommand{\rem}[1]{}

\newfont{\subsubsecfnt}{ptmri8t at 11pt}
\renewcommand{\subparagraph}[1]{\smallskip{\subsubsecfnt #1.}}

\newcommand{\eqnref}[1]{\hyperref[#1]{{(\ref*{#1})}}}
\newcommand{\thmref}[1]{\hyperref[#1]{{Theorem~\ref*{#1}}}}
\newcommand{\lemref}[1]{\hyperref[#1]{{Lemma~\ref*{#1}}}}
\newcommand{\corref}[1]{\hyperref[#1]{{Corollary~\ref*{#1}}}}
\newcommand{\defref}[1]{\hyperref[#1]{{Definition~\ref*{#1}}}}
\newcommand{\secref}[1]{\hyperref[#1]{{Sec.~\ref*{#1}}}}
\newcommand{\figref}[1]{\hyperref[#1]{{Fig.~\ref*{#1}}}}  
\newcommand{\figureref}[1]{\hyperref[#1]{{Figure~\ref*{#1}}}}  
\newcommand{\tabref}[1]{\hyperref[#1]{{Table~\ref*{#1}}}}
\newcommand{\remref}[1]{\hyperref[#1]{{Remark~\ref*{#1}}}}
\newcommand{\appref}[1]{\hyperref[#1]{{Appendix~\ref*{#1}}}}
\newcommand{\claimref}[1]{\hyperref[#1]{{Claim~\ref*{#1}}}}
\newcommand{\factref}[1]{\hyperref[#1]{{Fact~\ref*{#1}}}}
\newcommand{\propref}[1]{\hyperref[#1]{{Proposition~\ref*{#1}}}}
\newcommand{\exampleref}[1]{\hyperref[#1]{{Example~\ref*{#1}}}}
\newcommand{\conjref}[1]{\hyperref[#1]{{Conjecture~\ref*{#1}}}}

\allowdisplaybreaks[1]

\def\COLOR{}
\ifdefined\COLOR

\else

\fi

\definecolor{Cayenne}{rgb}{0.5,0,0}
\definecolor{Midnight}{rgb}{0,0,0.5}
\definecolor{Plum}{rgb}{0.5,0,0.5}
\definecolor{Teal}{rgb}{0,0.5,0.5}
\definecolor{Clover}{rgb}{0,0.5,0}
\definecolor{Maroon}{rgb}{0.5,0,0.25}
\definecolor{Ocean}{rgb}{0,0.25,0.5}
\definecolor{Tangerine}{rgb}{1,0.5,0}
\definecolor{Strawberry}{rgb}{1,0,0.5}
\definecolor{Fern}{rgb}{0.25,0.5,0}
\definecolor{Aqua}{rgb}{0,0.5,1}
\definecolor{Moss}{rgb}{0,0.5,0.25}
\definecolor{Mocha}{rgb}{0.5,0.25,0}
\definecolor{Lemon}{rgb}{1,1,0}
\definecolor{Asparagus}{rgb}{0.5,0.5,0}
\definecolor{Grape}{rgb}{0.5,0,1}
\definecolor{Iron}{rgb}{.3,.3,.3}
\definecolor{Steel}{rgb}{.4,.4,.4}
\definecolor{Purple}{rgb}{.5,0,.5}